# Data-Driven Transient Stability Assessment of Power Systems with a Novel GHM-Enhanced CatBoost Algorithm

Zheheng Wang
(Nanyang Institute of Technology, Nanyang City, Henan Province 473000, China)

**ABSTRACT:** This study introduces an advanced transient stability assessment (TSA) method for power systems, addressing the challenges of sample class imbalance and data noise through a novel CatBoost algorithm framework. By implementing a Gradient Harmonizing Mechanism (GHM), this method adjusts the gradient norm distribution across samples by incorporating a coordination parameter for each, thus optimizing the gradient weights for various sample types. This enhancement enables more effective training of the CatBoost algorithm, reducing the negative impacts of class imbalance and noise, and enhancing algorithmic performance. Additionally, the feature importance functionality of the CatBoost framework guides the placement of phasor measurement units, promoting economical operation of the power system. Numerical results from the New England 10-machine 39-bus system demonstrate the superior versatility, reduced application cost, and lower maintenance expenses of the proposed method compared to existing techniques.

## 0 Introduction

Damage to transient stability can precipitate cascading failures, potentially leading to extensive power outages with significant economic costs and negative societal impacts [1]. Timely and accurate detection of transient stability issues is essential for facilitating proactive management and decision-making, thus preventing the collapse of system stability and reducing economic losses [2]. The development of rapid and precise transient stability assessment (TSA) methods holds substantial practical significance.

Presently, research into TSA is bifurcated into two distinct streams: mechanistic analysis and machine learning approaches. Mechanistic analysis in TSA typically encompasses methods such as time-domain simulations and direct techniques, which include the extended equal-area method and the transient energy function method [3-4]. Despite their established utility, these traditional approaches struggle to keep pace with the increasingly intricate operation modes prompted by the integration of large-scale renewable energy sources, the deployment of high-voltage direct current (HVDC) transmission lines, and the extensive use of power electronic devices. These factors complicate the modern electric power system, diminishing the timeliness and effectiveness of conventional mechanistic analyses [5].

As a result, machine learning has emerged as a promising alternative, offering novel solutions to the challenges of TSA. The development of new-generation smart grids has facilitated the widespread adoption of Wide Area Measurement Systems (WAMS) and synchronized Phasor Measurement Units (PMUs), significantly improving the accessibility of power system data [6]. This enhancement has contributed to the increasing prevalence of machine learning techniques in TSA. Recent research has seen the application of various machine learning algorithms such as Convolutional Neural Networks (CNNs) [8], Deep Residual Networks (ResNets) [9], LightGBM [10-11], and XGBoost [7] to address and refine TSA processes.

In recent years, the field of power system TSA has seen significant advancements through the integration of machine learning techniques and novel computational frameworks [12-16]. A diverse array of studies underscores this progress, each focusing on different aspects of TSA to enhance reliability and accuracy in real-time applications [17]. The necessity for precise and efficient data handling methods is highlighted to address the sheer volume and variety of data typical in modern power systems [18-20]. Their reviews call attention to the crucial role of data-driven approaches for short-term voltage stability, which are essential for adapting to the dynamic nature of power grids. Moreover, the evolution of specific machine learning models tailored to TSA tasks is well documented [21-25]. For instance, Liu et al. and Li et al. have refined the application of deep learning models like Gradient Harmonizing Mechanism (GHM)-CatBoost and transfer learning to enhance the predictive accuracy and generalizability of TSA [26-28]. These models not only address the class imbalance and noise in training datasets but also ensure that the systems are adaptable to varied operational conditions. The integration of real-time data analytics further demonstrates the shift towards deploying advanced neural networks and other sophisticated machine learning algorithms that can process synchrophasor data effectively [29-31]. These developments not only enhance the real-time assessment capabilities but also ensure that the systems can handle the complexities of modern power grids with high penetrations of renewable energy sources. Overall, the collective research efforts, spanning from foundational reviews to advanced algorithmic implementations, illustrate a vibrant and rapidly evolving landscape in power system TSA [32-34]. This ongoing research not only addresses current challenges but also sets the groundwork for future innovations that will continue to enhance the stability and reliability of power systems worldwide.

In practice, the inherent robustness of power systems typically allows for recovery from large disturbances, making transient instability a relatively infrequent occurrence [35-38]. This results in a natural imbalance between the numbers of stable and unstable samples, which can complicate model training. If unchecked, this imbalance can lead to the misclassification of unstable states, with potentially grave consequences [39-41]. Recent studies have explored various methods to correct this imbalance, including the use of improved Conditional Generative Adversarial Networks (CGANs) for data augmentation and the application of sample weighting based on inverse category frequency. However, these methods often do not fundamentally enhance model performance and can lead to overfitting [42, 43]. Other efforts have employed the focal loss function to mitigate sample category imbalances, but its complex hyperparameter tuning and overemphasis on difficult samples can also lead to model overfitting [44, 45].

To address these challenges comprehensively, this paper proposes a TSA methodology based on an innovative CatBoost algorithm framework that incorporates a GHM. This novel approach adjusts the gradient weights of the loss function across stable and noisy samples during the training of the GHM-CatBoost model, effectively mitigating problems related to sample category imbalance and data noise. Moreover, the Feature Importance (FI) function within the GHM-CatBoost framework offers valuable guidance for optimal PMU device placement. The effectiveness of this methodology is validated on a 39-node system of 10 machines in New England, showcasing its broad applicability and robustness in real-world settings.

## 1 The GHM-CatBoost algorithmic framework

### 1.1 Principle of CatBoost Algorithm

CatBoost, an open-source machine learning library released by the Russian technology powerhouse Yandex in 2017, is part of the gradient boosting decision tree (GBDT) family [47]. This framework utilizes symmetric trees, also known as oblivious trees, as its foundational learning model, and is renowned for its robust support for categorical features and high predictive accuracy. Additionally, CatBoost addresses common challenges in machine learning such as gradient bias and prediction shifts, thereby minimizing overfitting. This enhancement significantly improves both the precision and the generalizability of the algorithm.

**1) Iterative process of CatBoost algorithm**

CatBoost is constructed using a symmetric tree as the base learner and a recursive division of the entire feature space [24]. Assuming that CatBoost divides the entire feature space into $J$ disjoint regions, and the predicted value of each region is $b_j$, the decision tree $h$ can be expressed as:

$$h(\boldsymbol{x}) = \sum_{j=1}^{J} b_j \mathrm{I}_{\{x \in R_j\}} \quad (1)$$

Where: $\mathrm{I}_{\{x \in R_j\}}$ is the indicator function, defined as:

$$\mathrm{I}_{\{x \in R_j\}} = \begin{cases} 1 & \text{if } \boldsymbol{x} \in R_j \\ 0 & \text{otherwise} \end{cases} \quad (2)$$

CatBoost generates a base learner model in each round of iterations and minimizes the loss function for the current round of iterations. Assuming that the strong learner obtained in the last round of iterations is $F^{m-1}(\boldsymbol{x})$, and the loss function is $L(y,F^{m-1}(\boldsymbol{x}))$, the objective function $h^m$ for the current round of iterations is:

$$h^m = \underset{h\in H}{\arg\min}\, EL(y,F^{m-1}(\boldsymbol{x})+h(\boldsymbol{x})) \quad (3)$$

The negative gradient of the loss function is then used to fit an approximation of the loss for each round, and the negative gradient $-g^m(\boldsymbol{x},y)$ of the loss function for the last iteration is:

$$-g^m(\boldsymbol{x},y) = \frac{\partial L(y,F^{m-1}(\boldsymbol{x}))}{\partial F^{m-1}(\boldsymbol{x})} \quad (4)$$

Apply least squares estimation and approximate fit the objective function $h^m$ as:

$$h^m = \underset{h\in H}{\arg\min}\, E(-g^m(\boldsymbol{x},y)-h(\boldsymbol{x}))^2 \quad (5)$$

Finally, the strong learner for this iteration is obtained as:

$$F(\boldsymbol{x})^m = F(\boldsymbol{x})^{m-1} + \mu h^m \quad (6)$$

Where: $\mu$ is the learning rate, i.e., the model update step.

**2）Problems with prediction offsets**

In each iteration of the traditional GDBT framework, the loss function uses the same training samples to compute the gradient of the current model and fits the loss approximation for the current iteration based on this gradient. However, literature [16] and literature [17] point out that the gradient distribution $g^m(\boldsymbol{x}_k,\boldsymbol{y}_k)|\boldsymbol{x}_k$ computed using the same training samples is biased compared to the true distribution $g^m(\boldsymbol{x},\boldsymbol{y})|\boldsymbol{x}$ of the gradients in the data space, such that the objective function obtained from the approximate fitting according to Eq. (5) deviates from the definition in Eq. (3), and the gradient bias in turn causes prediction bias and affects the accuracy and generalization ability of the final model.

CatBoost overcomes the prediction bias problem by using a sort boosting method for unbiased computation of the gradient step in each iteration. The principle of the sort boosting method is as follows: for each sample $x_i$, CatBoost trains a separate model $M_i$ using a training set that does not contain sample $x_i$, and uses model $M_i$ to estimate the gradient for sample $x_i$. This gradient is then used to fit the base learner model for the current round of iterations.

1.2 GHM-CatBoost Principle

The CatBoost algorithm of the paper uses cross entropy (cross entropy) loss function as shown in equation (7).

$$L_{CE}(p,y) = \begin{cases} -\log p & y=1 \\ -\log(1-p) & y=0 \end{cases} \quad (7)$$

where $p$ is the probability that the model predicts transient stabilization for that sample, and $y$ is the corresponding categorical index for that sample. The gradient of the cross-entropy loss function for sample $\boldsymbol{x}$ is:

$$\frac{\partial L_{CE}}{\partial \boldsymbol{x}} = \begin{cases} p-1 & y=1 \\ p & y=0 \end{cases} \\ = p-y \quad (8)$$

The gradient modulus is defined as:

$$g = \left|\frac{\partial L_{CE}}{\partial \boldsymbol{x}}\right| = |p-y| = \begin{cases} 1-p & y=1 \\ p & y=0 \end{cases} \quad (9)$$

The value of the gradient modulus $g$ ranges from 0 to 1. If the value of $g$ is larger, the sample is considered to be more difficult to classify. Thus, the sample category imbalance problem and the data noise problem are attributed to the imbalance of the gradient mode length distribution of the samples (i.e., the imbalance of the distribution of the hard-to-split samples and the easy-to-split samples). Based on the gradient mode length distribution of the samples, the GHM is proposed to attach an adjustment parameter to each sample, thus balancing the gradient weights of each category of samples.

First, the interval over which the modulus length of the gradient is taken is divided into $Z$ unit regions, each of length $\varepsilon = 1/Z$. Thus, the concept of gradient density is defined:

$$GD(g) = \frac{R_{ind}(g)}{\varepsilon} = R_{ind}(g)Z \quad (10)$$

Where: $R_{ind}(g)$ is the number of samples distributed in the unit area; therefore, the coordination parameter for each sample is set as:

$$\beta_i = \frac{n}{GD(g_i)} \quad (11)$$

Where: $n$ is the number of samples. From equation (11), it can be seen that the weight of samples with high gradient density will be decreased and the weight of samples with low gradient density will be increased. The coordination parameter will be used to correct the cross-entropy loss function:

$$L_{GHM}(p_i, y_i) = \frac{1}{n}\sum_{i=1}^{n}\beta_i L_{CE}(p_i, y_i) = \sum_{i=1}^{n}\frac{L_{CE}(p_i, y_i)}{GD(g_i)} \quad (12)$$

Among the offline samples, the largest number is the easy-to-split transient stable samples, followed by the especially difficult-to-split samples, which may have some noise points or outliers [15]. Applying $L_{GHM}$ to CatBoost will reduce the weights of easy-to-split transient stable samples and especially difficult-to-split samples, which makes the CatBoost model focus more on more effective normal difficult samples, and thus improve the model performance.

## 2 TSA based on the GHM-CatBoost algorithmic framework

### 2.1 Database generation

The PSS/E software is utilized for simulating various system operations, with Python and MATLAB programs integrated to automate the collection of simulation data. To mirror the system's diverse operational modes accurately, load fluctuations are randomly varied between 70% and 130% of the reference load value. Simultaneously, generator outputs are adjusted to maintain power equilibrium, ensuring that the voltage at each node stays within a stable range of 0.95 to 1.05 p.u. During the simulation, a three-phase short-circuit fault is introduced along 10% to 90% of the transmission line, with fault durations set between 0.1 and 0.3 seconds. The entire simulation runs for a duration of 10 seconds. The assessment of transient stability is conducted by analyzing the power angle values of each generator, employing the following Transient Stability Index (TSI) as a critical metric for evaluation [18,24]:

$$\text{TSI} = \frac{360° - \delta_{\max}}{360° + \delta_{\max}} \quad (13)$$

where $\delta_{\max}$ is the maximum power angle difference between any two generators after the end of simulation. If TSI>0, the system is judged to be transiently stable, which is indicated by “1”; the system is judged to be transiently unstable, which is indicated by “0”.

Referring to the research results in [26] and verified by experiments, it is finally determined to take the voltage magnitude/phase angle of each node at the instant of fault removal and the active/reactive power of each transmission line as the input features. Then, the offline samples are constructed into matrix form:

$$\boldsymbol{D} = \begin{bmatrix} V_{1,1}\cdots V_{1,a} & \theta_{1,1}\cdots\theta_{1,a} & P_{1,1}\cdots P_{1,b} & Q_{1,1}\cdots Q_{1,b}, & y_1 \\ V_{2,1}\cdots V_{2,a} & \theta_{2,1}\cdots\theta_{2,a} & P_{2,1}\cdots P_{2,b} & Q_{2,1}\cdots Q_{2,b}, & y_2 \\ & & \cdots\cdots & & \\ V_{n,1}\cdots V_{n,a} & \theta_{n,1}\cdots\theta_{n,a} & P_{n,1}\cdots P_{n,b} & Q_{n,1}\cdots Q_{n,b}, & y_n \end{bmatrix} \quad (14)$$

where $n$ is the number of samples, $a$ is the number of system nodes, and $b$ is the number of transmission lines; $V_{i,j}$ and $\theta_{i,j}$ ($i$=1,2,…,$n$; $j$=1,2,…,$a$) are the voltage magnitude and phase angle of the $i$ th sample of node $j$; $P_{i,j}$ and $Q_{i,j}$ ($i$=1,2,…,$n$; $j$=1,2,…,$b$) are the active power and reactive power of the $i$ th sample of line $j$; $y_i$ ($i$=1,2,…,$n$) is the TSI index of sample $i$.

### 2.2 Performance evaluation indicators

Given the imbalanced distribution of sample categories, the overall accuracy rate may not fully capture the model's true performance. Consequently, the confusion matrix—a fundamental tool in machine learning—is employed to provide a more nuanced evaluation. The model's effectiveness is measured using three key metrics: the Accuracy Rate (ACC), the False Alarm Rate (FAR), and the Rejection Rate (PRR). These metrics collectively help to assess the strengths and weaknesses of the TSA model, offering a comprehensive overview of its performance in different scenarios.. The confusion matrix of TSA is shown in Tab. 1.

Tab.1 Confusion matrix

| Items | Real situations | |
|---|---|---|
| | Stability ($y$=1) | Instability ($y$=0) |
| Stability ($y$=1) | TP | FP |
| Instability ($y$=0) | FN | TN |

The accuracy rate ACC, false alarm rate FAR and truth rejection rate FRR are denoted as [22]:

$$\mathrm{ACC}=\frac{\mathrm{TP}+\mathrm{TN}}{\mathrm{TP}+\mathrm{TN}+\mathrm{FP}+\mathrm{FN}} \tag{15}$$

$$\mathrm{FAR}=\frac{\mathrm{FP}}{\mathrm{FP}+\mathrm{TN}} \tag{16}$$

$$\mathrm{FRR}=\frac{\mathrm{FN}}{\mathrm{TP}+\mathrm{FN}} \tag{17}$$

According to the above formula, it can be seen that: the accuracy rate ACC is the proportion of correctly classified samples among all samples, and can be used as a global indicator; the false alarm rate FAR is the proportion of destabilized samples identified as stable samples, which indicates the classification performance of the model for the destabilized samples; and the rejection rate FRR is the proportion of stabilized samples identified as destabilized samples, which indicates the classification performance of the model for the stabilized samples.

**2.3 Implementation Processes**

Fig.1 illustrates the process for implementing TSA. Initially, a subset of the offline samples is designated to create a training set, while another subset forms the test set. This training set is then processed through the GHM-CatBoost algorithm to train the model, establishing a connection between the input features and the corresponding Transient Stability Index (TSI) metrics, culminating in the development of the GHM-CatBoost-driven TSA model. Subsequently, the TSA model's efficacy is evaluated using the test set.

Once the TSA model achieves satisfactory predictive accuracy, it can be deployed for online TSA applications within the power system. This begins with the real-time collection of system operation data immediately after fault clearance using PMUs. The relevant input features are extracted as this data arrives at the dispatch center. The trained GHM-CatBoost model is then applied to predict transient stability conditions online.

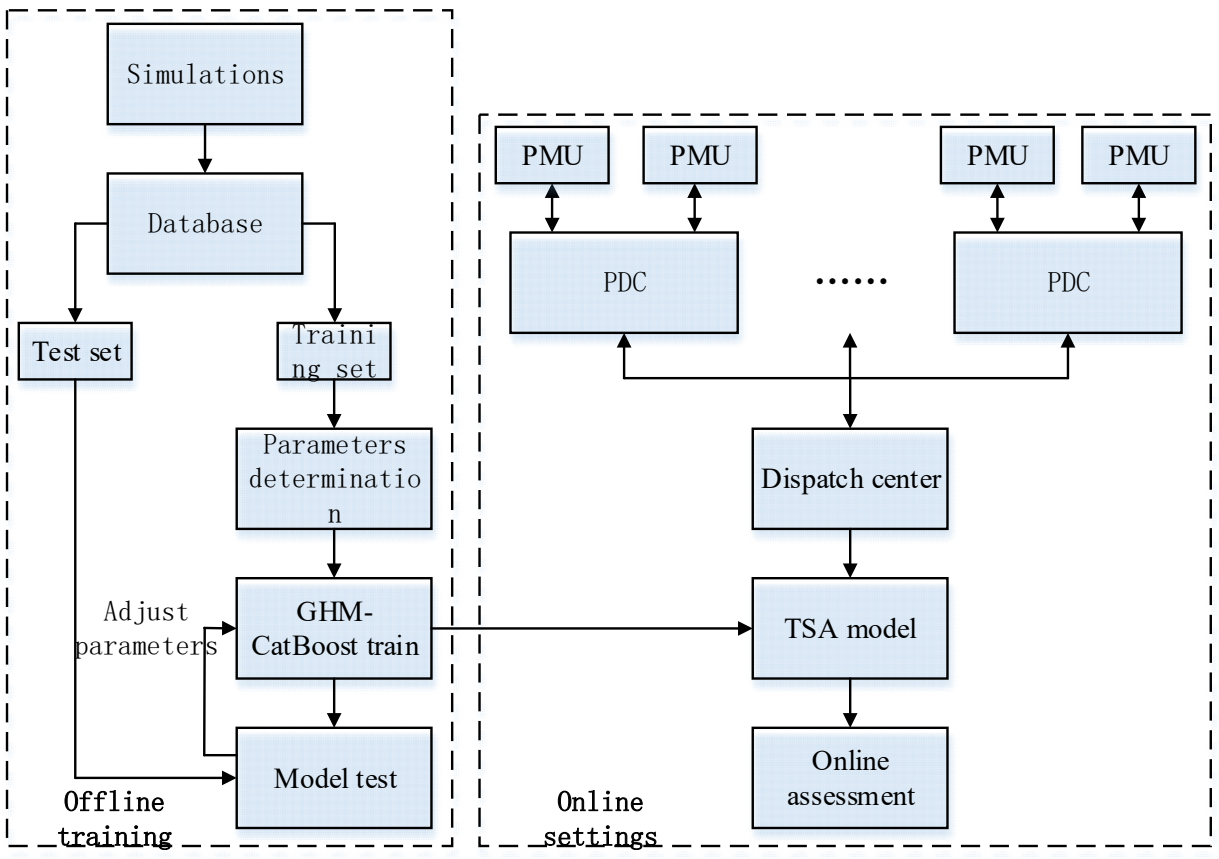

Fig.1 Flowchart of TSA

**3 Case Study**

The efficacy of the TSA method outlined in the paper is demonstrated using the New England 10-machine 39-node system. Figure 2 displays the system's topology, which includes 39 nodes, 10 generators, and 46 transmission lines [17]. For the purposes of these tests, it is presumed that PMUs are installed at every node. All evaluations are conducted on a computer equipped with an Intel Core i5 processor and 8 GB of RAM. Utilizing the simulation approach described in Section 2.1, a total of 5,897 samples were generated, comprising 4,809 stabilized samples and 1,088 destabilized samples.

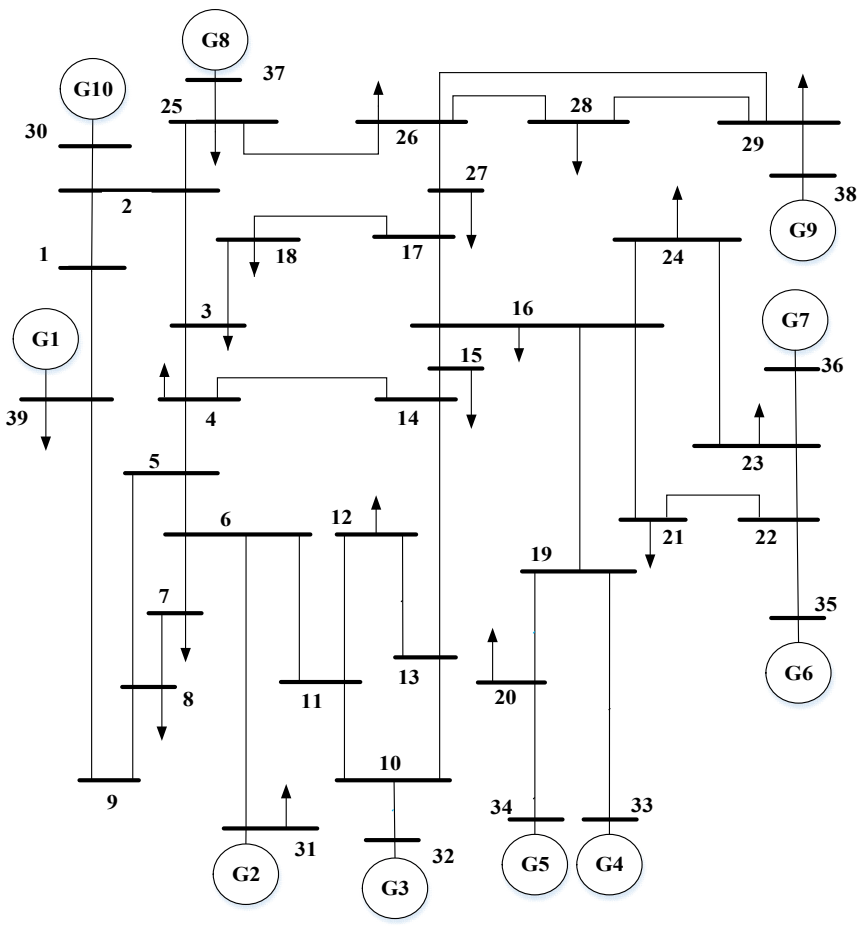

Fig.2 Topology of New England 10 machine 39 bus system

**3.1 Performance analysis of different algorithms**

**1）Prediction accuracy and training time**

80% of the samples were used for training and 20% of the samples were used for testing, and ResNet, XGBoost, LightGBM, CatBoost and GHM-CatBoost were used for TSA, respectively, and the 5-fold cross-validation method was used to repeat the experiment, and the average of the five tests was taken to verify the effectiveness of GHM-CatBoost. Among them, the various classification algorithms are built under the platform of Python open source machine learning library Scikit-learn [20], and the parameters are set as default parameters. The prediction accuracy and training time of different algorithms are shown in Tab. 2.

Tab.2 Accuracy rate of different classifiers

| Algorithms | ACC% | FAR% | FRR% | 训练时间/秒 |
|---|---|---|---|---|
| ResNet | 98.06% | 3.12% | 1.69% | 68.36 |
| XGBoost | 97.86% | 3.57% | 1.77% | 22.65 |
| LightGBM | 98.10% | 3.08% | 1.66% | 9.16 |
| CatBoost | 98.23% | 2.75% | 1.56% | 16.55 |
| GHM-CatBoost | 98.71% | 1.38% | 1.28% | 18.42 |

**2）Noise Resistance**

During the collection and transmission of power data, fluctuations in the communication network, equipment malfunctions, and other factors can introduce noise into the data. As per the IEEE C37.118 standard, the measurement error for the phase volume of a PMU should not exceed 1% [21]. To evaluate the resilience of different algorithms against noise, various percentages of Gaussian white noise (with a mean of zero and a standard deviation of one) are incorporated into the sample set. These modified samples are then processed using distinct algorithms to assess their anti-noise capabilities. Figures 3 and 4 illustrate the test results obtained at various noise levels.

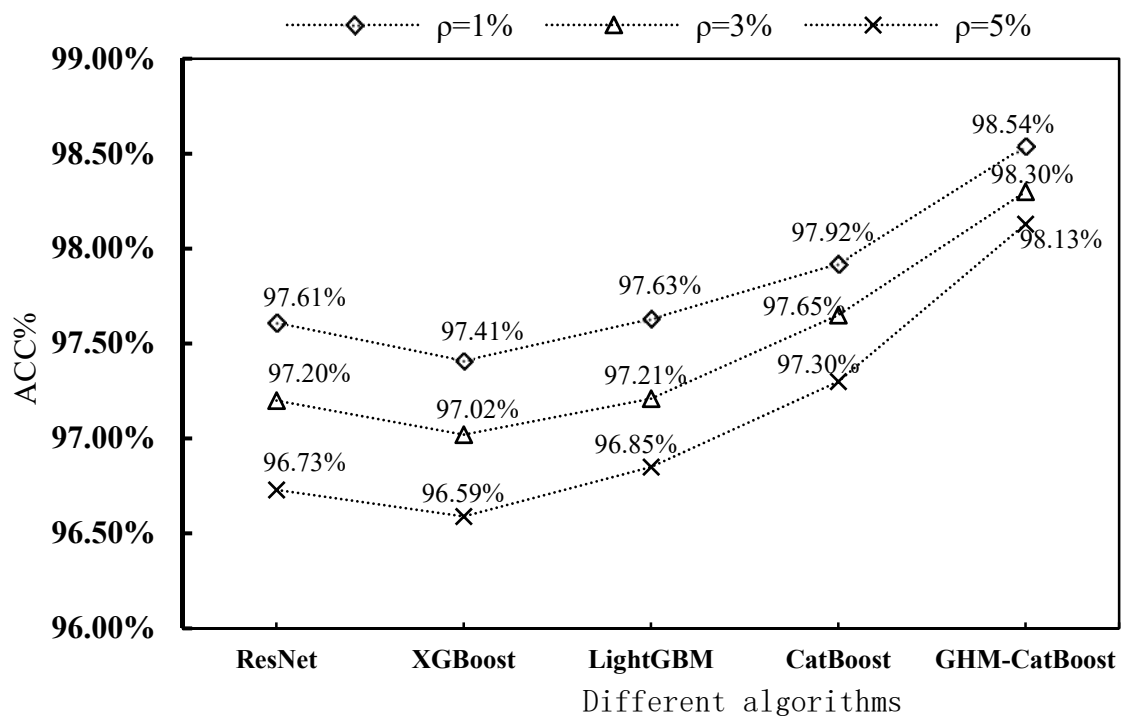

Fig. 3 ACC% of each algorithm under different noise levels

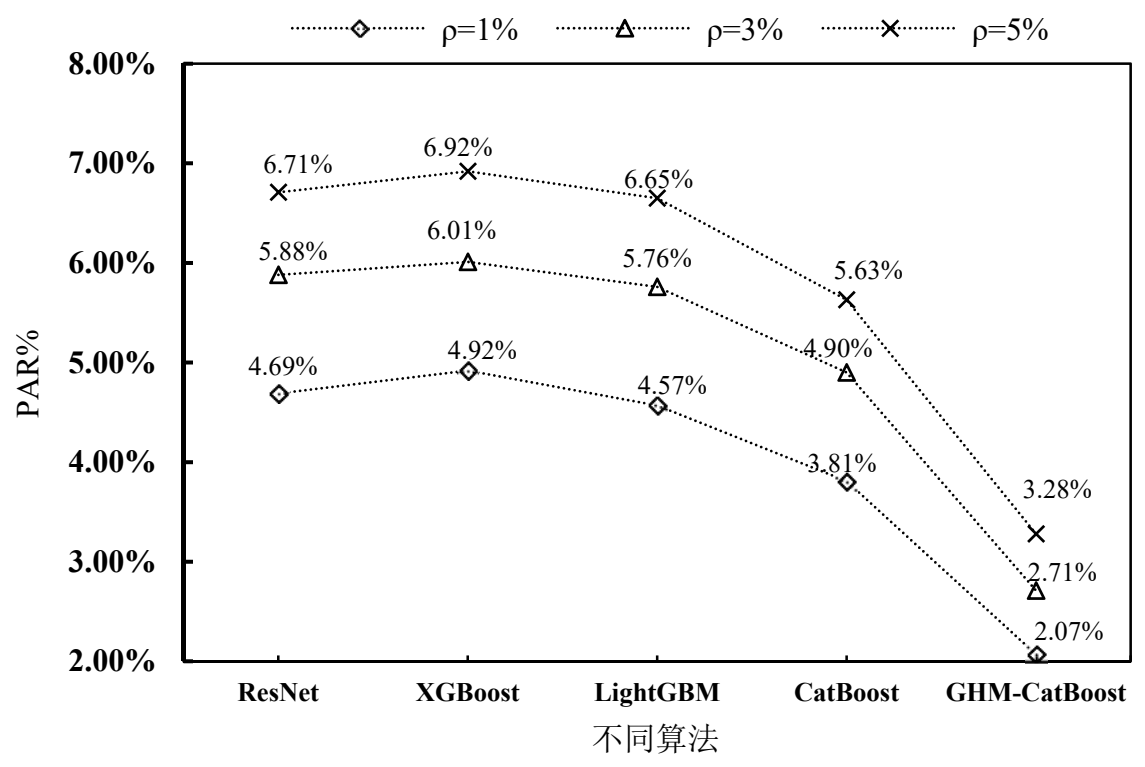

Fig.4 ACC% of each algorithm under different noise levels

The analysis of different algorithms based on the test results is detailed as follows:

**CatBoost, XGBoost, and LightGBM**: All three algorithms are contemporary implementations within the GBDT framework. CatBoost demonstrates slightly higher accuracy than LightGBM in the initial sample set, while XGBoost shows slightly lower accuracy. LightGBM excels in computational speed, but CatBoost, utilizing a symmetric tree as its base learner, combines speed with enhanced robustness. Moreover, CatBoost's unique sort boosting method for tree model construction effectively resolves the issue of prediction bias, giving it superior generalization capabilities compared to both XGBoost and LightGBM. This makes CatBoost particularly robust and practical, showcasing its broad applicability and efficiency.

**CatBoost and ResNet**: ResNet, a deep learning model, excels in processing high-dimensional data like images, speech, and text by capturing spatio-temporal relationships. However, CatBoost is more adept at handling tabular data. In tests, CatBoost not only outperforms ResNet in accuracy but also requires significantly less training time. CatBoost also surpasses ResNet in anti-noise performance, and its model construction demands fewer training samples and involves less intricate parameter tuning than ResNet. Consequently, CatBoost offers lower operational and maintenance costs.

**GHM-CatBoost**: In the original sample set, GHM-CatBoost shows a 0.48% improvement in accuracy over CatBoost, along with a 1.37% reduction in the False Alarm Rate and a 0.28% reduction in the Rejection Rate. The introduction of the GHM, which computes the gradient modulus length for each sample, slightly increases training times by 1.87 seconds. Furthermore, the inclusion of GHM enhances the model’s resistance to noisy data, thus bolstering the robustness of the CatBoost framework. These results underscore the effectiveness of the GHM-CatBoost in enhancing the model's resilience and performance.

### 3.2 Model performance analysis under unbalanced samples

In the sample set, 4000 samples are selected as the training set and different ratios of stabilized and destabilized samples are set, and another 1897 samples are used as the test set. Then, TSA is performed on CatBoost and GHM-CatBoost respectively to verify the effectiveness of GHM, and the test results are shown in Fig. 4.

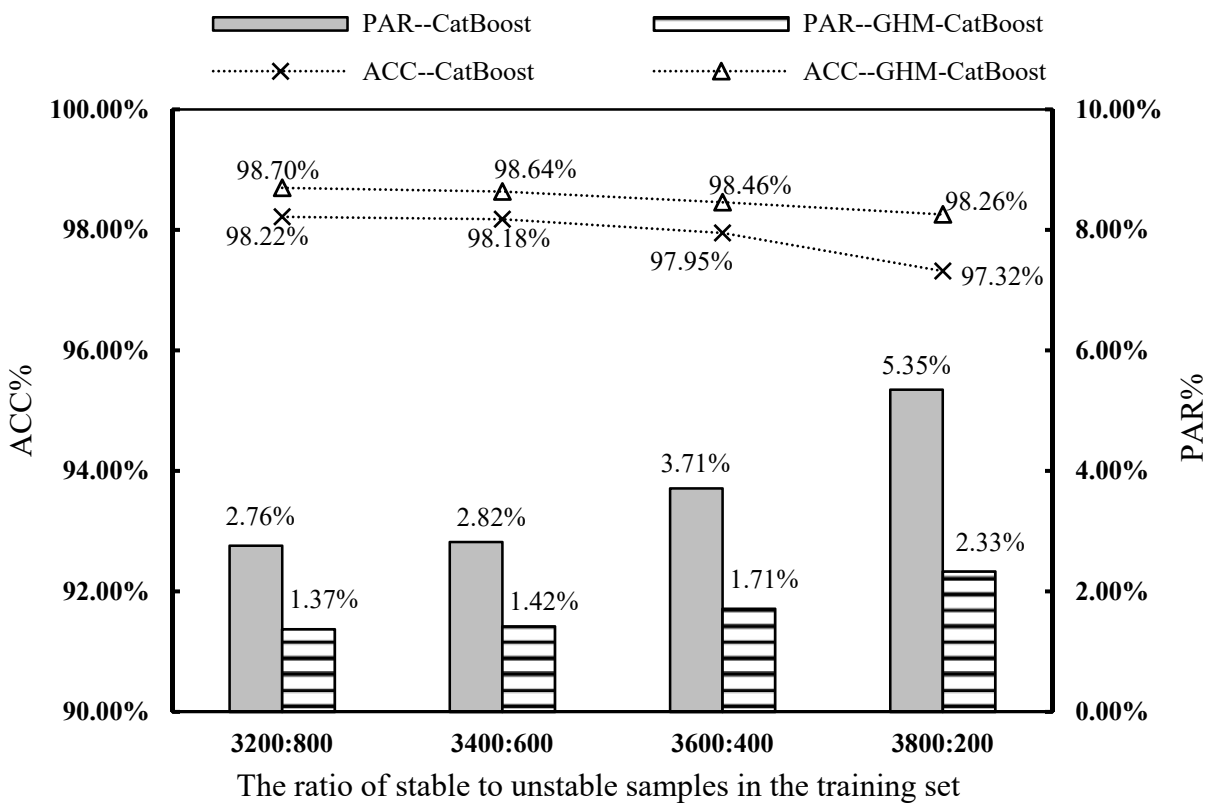

Fig. 5 Test results under different proportions of stable samples and unstable samples

The above test results show that the more severe the sample category imbalance in the training samples, the more frequently the classifier will have misclassification of the destabilized samples, this is because the total loss of the classifier predicting all the destabilized samples to be stabilized samples will also be small. In addition, as the ratio of stable to destabilized samples increases, GHM-CatBoost is less affected than CatBoost because the introduction of GHM assigns different weights to each sample, which balances the gradient weights between the stable and destabilized samples, and thus mitigates the sample category imbalance problem. Therefore, the introduction of GHM helps to guide the training of a more reliable TSA model, which also illustrates the effectiveness of GHM.

### 3.3 Analysis of PMU deployment

In practice, it is difficult to realize the full coverage of PMU devices in the whole network due to the price of PMU devices as well as the cost of communication networks [24]. Therefore, it is important to realize fast and accurate TSA with limited PMU configuration. The paper calculates the importance score of each feature based on the feature importance function of the GHM-CatBoost algorithm framework. Among them, the importance scores of the overall features are shown in Fig. 6. In addition, some of the features with the top importance scores are shown in Tab. 3.

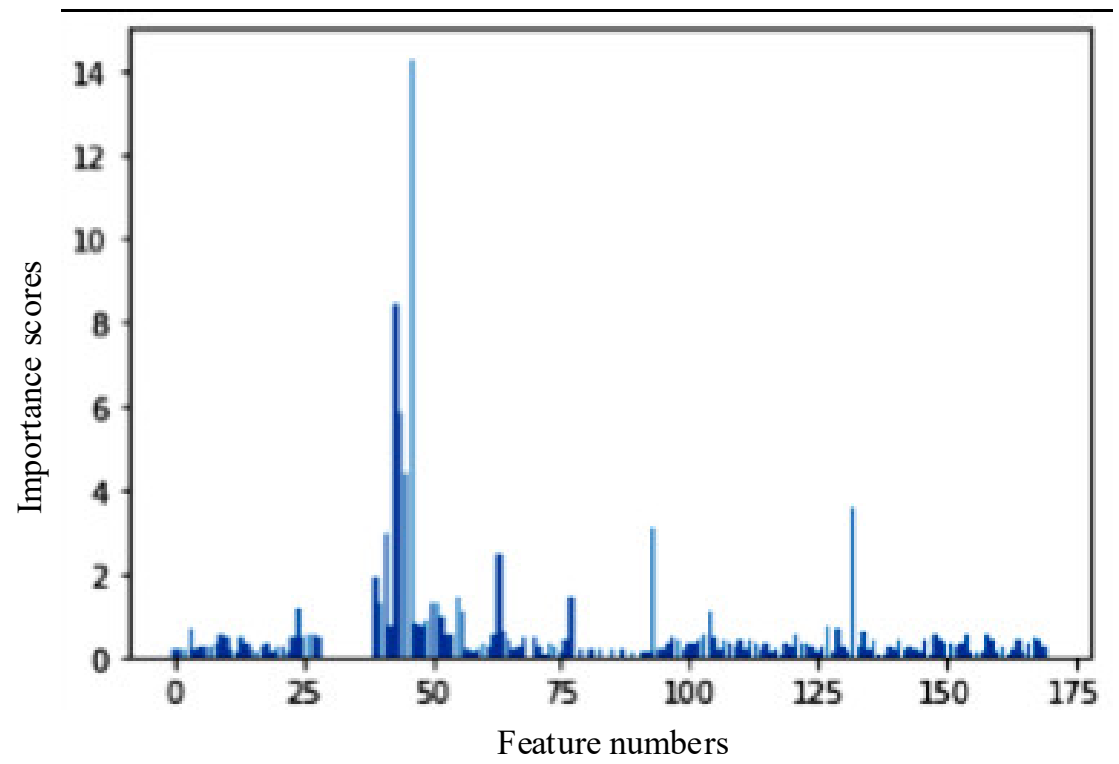

Fig. 6 The importance score of overall features

Tab.3 Some features of priority score ranking

| 排名 | Feature no | Features | Scores |
|---|---|---|---|
| 1 | 46 | Voltage phase angle of node 8 | 14.32 |
| 2 | 43 | Voltage phase angle of node 5 | 8.42 |
| 3 | 44 | Voltage phase angle of node 6 | 5.87 |
| 4 | 45 | Voltage phase angle of node 7 | 4.41 |
| 5 | 132 | Reactive power of line 4-14 | 3.60 |
| 6 | 93 | Active power of line 8-9 | 3.10 |
| 7 | 41 | Voltage phase angle of node 3 | 2.98 |
| 8 | 63 | Voltage phase angle of node 25 | 2.47 |
| 9 | 39 | Voltage phase angle of node 1 | 1.91 |
| …… | …… | …… | …… |

Based on the feature importance scores of the GHM-CatBoost outputs, it can be seen that the correlation between the node voltage phase angle and the transient stability condition is higher than the other input features. If the PMU device is installed on a node, the voltage magnitude/phase angle of the node, and the active/reactive power of the connected line can be collected online, and according to the feature importance score situation of the overall features, the PMU device is installed on the node with high importance, and the PMU deployment scheme as shown in Tab. 4 is designed, then the results of the model evaluation under different schemes are shown in Tab. 5.

Tab.4 The layout scheme of PMU device

| Scheme | Counts | Installation positions（bus number） |
|---|---|---|
| 1 | 5 | 8、5、6、7、14 |
| 2 | 10 | 8、5、6、7、14、9、1、39、17、12 |
| 3 | 15 | 8、5、6、7、14、9、1、39、17、12、2、13、25、16、18 |
| 4 | 20 | 8、5、6、7、14、9、1、39、17、12、2、13、25、16、18、14、11、9、4、10 |

Tab.5 ACC% of the model under different PMU layout schemes

| Model | ACC% | | | |
|---|---|---|---|---|
| | Scheme 1 | Scheme 2 | Scheme 3 | Scheme 4 |
| GHM-CatBoost | 98.59% | 98.66% | 98.71% | 98.71% |

As shown in Table 5, the accuracy percentage of PMU deployment can achieve 98.59% with just 5 units installed; exceeding 10 installations approximates the ACC% achievable with full coverage, indicating redundancy in the original sample set. When applied to a large-scale power grid, reducing redundant input features based on their importance scores can significantly enhance the computational efficiency of the model. Thus, the GHM-enhanced CatBoost framework provides valuable guidance for PMU placement.

## 4 Conclusion

Addressing the challenges of sample category imbalance and data noise, this paper integrates GHM into the CatBoost framework to develop a refined TSA methodology. Analysis conducted on the New England 10-machine, 39-bus system yields several key insights:

1) CatBoost demonstrates superior generalization, cost-effectiveness, and lower maintenance requirements compared to ResNet, XGBoost, and LightGBM. The integration of GHM effectively mitigates issues related to sample imbalance and data noise.
2) The GHM loss function is dynamically adjusted based on the distribution of gradient modulus length, eliminating the need for manual parameter tuning and reducing training costs.
3) The feature importance functionality within the GHM-CatBoost framework not only aids in efficient PMU siting but also enhances grid operational economy by eliminating redundant features and improving computational efficiency.

The current study acknowledges the strengths of the CatBoost algorithm in handling categorical data but does not yet incorporate fault location and type as input features. Future research will aim to include these fault details as categorical inputs into the GHM-CatBoost framework to enhance the model's ability to detect and respond to sophisticated cyber threats [48-50], including false data injections [51-52]. Additionally, as cybersecurity threats to power systems increase, incorporating advanced security measures like federated learning within the TSA process is crucial [53-55]. This approach will help protect data privacy while enabling collaborative learning necessary for robust energy management [56-57]. The next phase of research will validate this TSA method using actual operational data from power systems, focusing on the accuracy of stability assessments and the effectiveness of data protection mechanisms. This future testing will assess the model's performance in real-world settings, ensuring it can handle the complexities of modern power systems securely and efficiently.